\documentclass[10pt,twocolumn,letterpaper]{article}

\usepackage[
  letterpaper,
  top=0.75in,
  bottom=0.80in,
  left=0.70in,
  right=0.70in,
  columnsep=0.25in
]{geometry}

\usepackage[T1]{fontenc}
\usepackage[utf8]{inputenc}
\usepackage{times}
\usepackage{helvet}
\usepackage{courier}

\usepackage{amsmath}
\usepackage{amssymb}
\usepackage{bm}

\usepackage{graphicx}
\usepackage{array}
\usepackage{multirow}
\usepackage{caption}

\usepackage{algorithm}
\usepackage{algorithmic}

\usepackage[round,authoryear]{natbib}
\usepackage[hyphens]{url}
\usepackage[hidelinks]{hyperref}

\usepackage{authblk}

\title{
What Makes a Fairness Gap Actionable?\\
Statistical Actionability for Responsible AI Deployment
}

\author[1]{Hairu Fan\thanks{Corresponding author: fan2h@cmich.edu}}
\author[2]{Shiyuan Wang}

\affil[1]{
Department of Statistics, Actuarial, and Data Sciences,\\
Central Michigan University, Mount Pleasant, MI, USA
}

\affil[2]{
Central Michigan University, Mount Pleasant, MI, USA
}

\date{}

\begin{document}

\maketitle

\begin{abstract}
Algorithmic fairness audits can detect disparities, but they do not determine when those disparities warrant intervention. Deployment decisions also depend on the reliability of the evidence, subgroup support, and deployment context. Existing fairness methods quantify disparities and uncertainty, yet provide limited guidance for translating accumulated evidence into action. We introduce Statistical Actionability, a statistical construct that recasts fairness deployment as an evidence-based decision problem. The framework integrates fairness evidence regarding disparity magnitude, statistical reliability, subgroup adequacy, and deployment context, and maps the resulting evidence state to one of four recommendations: mitigate, collect more data, monitor, or take no immediate action. In controlled simulations, Statistical Actionability achieved the lowest decision cost among representative baselines, reducing average decision cost by 19.2\% relative to gap-based intervention while simultaneously reducing both false alarms and missed bias. A calibrated deployment rule generalized across heterogeneous statistical environments, remaining within 2\% of the target oracle in four of five transportability regimes. Analyses of benchmark fairness audits further demonstrated that the framework distinguished audits with similar observed fairness gaps but different levels of uncertainty and subgroup support, yielding interpretable deployment recommendations. Statistical Actionability therefore establishes a statistical decision layer between fairness evaluation and deployment intervention, enabling responsible AI systems to act on accumulated evidence rather than disparity magnitude alone.
\end{abstract}

\section{Introduction}

Risk-based {AI} governance is increasingly asking organizations to justify how they identify, evaluate, and respond to harms in deployed systems: the {EU} {AI} Act mandates lifecycle risk management for high-risk systems, including the identification, estimation, evaluation, and mitigation of risks to health, safety, and fundamental rights \cite{eu2024aiact}, while the {NIST} {AI} Risk Management Framework similarly frames trustworthy deployment as an iterative cycle of identifying, measuring, managing, and monitoring risk, albeit voluntarily \cite{nist2023airmf}. Both create a practical demand for evidence-based deployment decisions: when an audit reveals a fairness disparity, organizations must decide whether the evidence is sufficient to justify corrective action, continued monitoring, additional data collection, or no immediate intervention.

This demand outpaces the technical infrastructure meant to support it. A decade of fairness research has produced well-defined metrics such as equal opportunity \cite{hardt2016equality}, consolidated into auditing toolkits such as {Fairlearn} \cite{weerts2023fairlearn} and synthesized in comprehensive surveys \cite{mehrabi2021survey}---yet none of this machinery answers the question that follows measurement. A toolkit can report a four-point disparity in true positive rates across two subgroups, perhaps with a confidence interval; it cannot say whether four points, on this sample, under this subgroup imbalance and this deployment cost, constitutes sufficient evidence for intervention.

In the absence of a principled answer, teams improvise in both directions: some retrain models in response to disparities that are primarily sampling noise, incurring cost without fairness gain, while others dismiss disparities that reflect persistent, consequential inequities, leaving harms unaddressed. Neither error is visible in the fairness metric itself, which is designed to report a disparity rather than adjudicate the sufficiency of evidence for acting on it.

We argue this is not a flaw in any specific metric, but a missing statistical layer between two problems the field has treated as one: fairness evaluation and fairness deployment. Recent work treats fairness auditing as an inferential exercise rather than a point estimate \cite{cherian2023statistical}, and practitioner studies document uncertainty about how to act on the disparities such audits surface \cite{madaio2022assessing}---but neither specifies when the resulting evidence clears the bar for action. We term this missing construct \emph{Statistical Actionability}: the degree to which accumulated statistical evidence justifies a fairness intervention under a specific deployment context. The distinction echoes one clinical practice makes routinely, between a biomarker and a diagnostic threshold: an elevated marker is a signal, not a prescription, and whether it warrants treatment depends on measurement precision, the weight of prior evidence, and the cost of acting versus waiting. Fairness gaps call for the same evidentiary reasoning, and the field currently lacks a formal way to supply it.

This paper develops such a framework and evaluates it empirically. Our contributions are as follows. \emph{Formalization}: we cast Statistical Actionability as an evidence-integration problem in which signal strength, statistical precision, evidence adequacy, and deployment context jointly determine whether the appropriate response is to mitigate, monitor, collect more data, or take no immediate action. \emph{Simulation evidence}: across controlled simulations, we show that observed fairness gaps, taken alone, are unreliable triggers for deployment intervention, producing both false alarms and missed interventions. \emph{Ablation analysis}: we isolate which components of the evidence-integration process drive actionability judgments, clarifying which forms of statistical evidence matter most. \emph{Calibration and transportability}: we show that the evidence thresholds our framework yields are not fixed constants, but are calibratable, context-dependent, and robust across several deployment settings while revealing important boundary conditions. \emph{Real-world deployment analysis}: applying the framework to established fairness benchmarks, we show that statistical evidence substantially reshapes deployment recommendations, reallocating many gap-triggered interventions to monitoring or additional data collection when the supporting evidence is insufficient for immediate mitigation.

Framed this way, responsible {AI} deployment is less a question of which fairness metric to compute than of how to integrate the evidence that metric provides---a distinction current governance frameworks increasingly require in practice but do not yet operationalize statistically.

\section{Related Work}

Responsible {AI} deployment sits at the intersection of three literatures that have largely developed in parallel: work that defines and measures fairness gaps, work that treats those measurements as statistical estimates subject to uncertainty, and work that studies how fairness considerations are operationalized inside real deployment pipelines. We review each before positioning our contribution relative to all three.

\subsection{From Fairness Metrics to Fairness Assessment Tools}

Fairness metrics are choices, not objective facts: the choice of metric---and the assumptions it encodes about which errors matter and to whom---shapes what a ``gap'' even means in a given deployment \cite{mitchell2021algorithmic}. \citeauthor{castelnovo2022clarification} \citeyear{castelnovo2022clarification} show that the landscape of fairness metrics is considerably more nuanced than the canonical definitions dominating applied work suggest, with different metrics supporting conflicting conclusions about the same model. Equal opportunity \cite{hardt2016equality}---the metric our empirical work builds on---asks whether true positive rates are equalized across groups.

What has changed in recent years is less the metrics themselves than their packaging. Toolkits such as {Fairlearn} \cite{weerts2023fairlearn} now compute a range of group fairness definitions across affected subpopulations and pair the resulting disparities with mitigation algorithms; {AIF360} \cite{bellamy2018aif360} and {Aequitas} \cite{saleiro2018aequitas} offer comparable functionality for auditors and practitioners. \citeauthor{mehrabi2021survey} \citeyear{mehrabi2021survey} synthesize this landscape of bias sources, definitions, and mitigation strategies into a survey that reflects how mature fairness assessment has become as a technical field.

These tools make disparities visible, but visibility is not actionability. A toolkit can report that a disparity exists and provide access to mitigation procedures, but it does not determine when that disparity is sufficiently supported by statistical evidence to justify intervention---the measurement layer, not the statistical decision layer, on which deployment depends.

\subsection{Statistical Uncertainty in Fairness Auditing}

A more recent line of work treats a fairness metric as an estimate rather than a settled fact. \citeauthor{cherian2023statistical} \citeyear{cherian2023statistical} reframe fairness auditing as statistical inference, casting subpopulation-level auditing as a multiple hypothesis testing problem with bootstrap-based simultaneous guarantees. \citeauthor{ji2020trust} \citeyear{ji2020trust} ask a related question from a Bayesian angle---how much a practitioner should trust a metric computed from limited or unlabeled data \cite{dimitrakakis2017bayesian}. Together, these establish a premise our framework inherits: a fairness gap is a random variable, not a fixed quantity.

The consequences are more disruptive than a confidence interval alone suggests. \citeauthor{barrainkua2023uncertainty} \citeyear{barrainkua2023uncertainty} show that fairness conclusions can flip under ordinary sampling fluctuation, while \citeauthor{khan2023fairness} \citeyear{khan2023fairness} ask whether estimator variance should be minimized or read as a signal of how much confidence an audit deserves. Together, these studies suggest that the instability we convert into evidentiary signal, rather than treat as noise, is a documented feature of fairness auditing, not an artifact of our setup.

A third strand quantifies this uncertainty more precisely---through calibrated confidence in fairness claims \cite{roy2023fairness} or formal statistical guarantees at the estimator level \cite{hu2023parametric}---but precision alone is not a decision rule. Fairness metrics are statistical estimates, not deterministic facts, yet most uncertainty-aware auditing stops at inference, certification, or confidence reporting; none specifies how uncertainty, evidence strength, subgroup adequacy, and deployment context should jointly determine whether to mitigate, monitor, collect more data, or take no action.

\subsection{Operationalizing Fairness in Deployment}

A separate, empirical literature asks what happens when fairness assessment meets an actual deployment pipeline. \citeauthor{madaio2022assessing} \citeyear{madaio2022assessing} find that disaggregated fairness evaluation routinely stalls on questions our framework formalizes: which metric to trust, which stakeholder groups warrant separate analysis, and how much data is enough to act on. \citeauthor{quinonero2023disentangling} \citeyear{quinonero2023disentangling} similarly argue, from LinkedIn's fairness operationalization, that what counts as actionable cannot be specified in the abstract but depends on product context and the population a system serves---consistent with the context dependence we build into Statistical Actionability directly.

This need for defensible deployment judgment has increasingly been codified into governance guidance. The {NIST} {AI} Risk Management Framework frames trustworthy deployment as a cycle of identifying, measuring, managing, and monitoring risk, but stops short of specifying how measured risk should be weighed against the cost or urgency of acting on it \cite{nist2023airmf}. Others target parts of this gap more narrowly: end-to-end lifecycle auditing \cite{raji2020closing}, causal inference for measurement under confounding \cite{enouen2023measuring}, and post-processing for counterfactual equalized odds in high-stakes instruments \cite{mishler2021fairness}---yet \citeauthor{bateni2022fairness} \citeyear{bateni2022fairness} are right that no single technical fix resolves this: the move from principle to practice is, unavoidably, a matter of human judgment about costs and targets.

\subsection{Positioning}

Table~\ref{tab:positioning} summarizes where existing streams of fairness research stop relative to the capabilities required for deployment actionability. The entries describe primary analytical emphasis rather than a binary checklist of individual papers.

\begin{table}[t]
\centering
\small
\setlength{\tabcolsep}{4pt}
\renewcommand{\arraystretch}{1.12}
\begin{tabular}{p{0.33\columnwidth} p{0.25\columnwidth} p{0.31\columnwidth}}
\hline
Need for actionability &
Prior emphasis &
Our framework \\
\hline
Observed gap &
Measurement &
Signal Evidence \\
Uncertainty &
Reliability &
Precision Evidence \\
Subgroup support &
Partial attention &
Adequacy Evidence \\
Deployment context &
Operational judgment &
Contextual Requirements \\
Action choice &
Limited formalization &
Action Set \\
\hline
\end{tabular}
\caption{Positioning Statistical Actionability relative to existing fairness research. The framework integrates measurement, uncertainty, subgroup evidence, and deployment context into deployment recommendations.}
\label{tab:positioning}
\end{table}

No existing literature, to our knowledge, closes the gap between detecting a disparity and deciding whether it warrants action. Section~\ref{sec:framework} formalizes that missing layer as Statistical Actionability, recasting fairness deployment as statistical decision-making under uncertainty and specifying the evidence dimensions, decision rule, and evaluation procedure this requires.

\section{Statistical Actionability Framework}
\label{sec:framework}
\subsection{Fairness Deployment as Statistical Decision-Making}
\label{sec:decision-framing}

Estimating a population-level disparity from finite data is a problem of statistical inference; deciding whether to retrain a model, monitor it further, collect more data, or maintain the current deployment is a problem of decision-making under uncertainty. Contemporary fairness pipelines often treat these as one problem, but they are not.

An observed fairness gap---whether demographic parity, equalized odds, or a difference in true positive rates---is a finite-sample estimate subject to sampling variability, subgroup imbalance, and measurement error, so identical observed gaps can correspond to substantially different underlying disparities. Deployment decisions also carry asymmetric consequences: intervening on a gap that is primarily sampling noise wastes engineering effort or degrades predictive performance, while failing to intervene on a persistent disparity leaves harms unaddressed.

Existing pipelines largely elide this distinction---toolkits transform outputs into fairness metrics, and the transition to deployment decisions is left to manually chosen thresholds or institutional convention. We argue this gap arises from a specific and correctable conflation: a fairness metric is treated as if it were a deployment decision, when it should instead be one input to such a decision.

\subsection{Statistical Actionability Principle}
\label{sec:sap}

The distinction between fairness evaluation and fairness deployment motivates the following principle.

\textbf{Statistical Actionability Principle (SAP).} Fairness intervention should be triggered when the accumulated statistical evidence is sufficient to justify intervention under the current deployment context, rather than solely according to the magnitude of an observed fairness gap.

Three implications follow. First, a fairness metric is evidence, not a verdict---it describes what has been observed, not what should be done about it. Second, because deployment risk rarely turns on a single statistic, evidence should be integrated rather than thresholded metric by metric. Third, how much evidence is ``enough'' depends on the deployment setting---the cost of intervention, the stakes of delay, the reversibility of action, and the tolerance for uncertainty---rather than being a universal constant.

The relevant question is therefore no longer only whether a fairness gap is observed, but whether the evidence supporting it is strong, precise, and adequate enough, in the current context, to justify intervention.

\subsection{Evidence Dimensions and Actionability Estimation}
\label{sec:evidence}

Unlike a fairness metric, Statistical Actionability is not directly observable from data. It is a latent deployment quantity---the degree to which accumulated statistical evidence supports intervention under a given deployment context---and should not be mistaken for another fairness definition alongside demographic parity, equalized odds, or equal opportunity; it operates at a higher inferential level.

We organize the supporting evidence into three complementary dimensions. \textbf{Signal Evidence} asks whether a meaningful disparity is likely to exist at all, capturing the strength of the statistical indication that the observed gap reflects a genuine difference rather than random variation. \textbf{Precision Evidence} asks how much confidence that indication deserves, given how much finite-sample uncertainty allows the signal to support a confident decision. \textbf{Adequacy Evidence} asks whether enough relevant subgroup information has accumulated---even a strong apparent signal can make immediate intervention premature if subgroup support is sparse.

Alongside these sits \textbf{Deployment Context}: not itself a form of statistical evidence, but the cost, stakes, reversibility, regulatory requirements, and institutional tolerance that govern how much evidence is required.

Let \(A\) denote Statistical Actionability, \(E_s\), \(E_p\), \(E_a\) the Signal, Precision, and Adequacy dimensions, and \(C\) Deployment Context. Conceptually,
\[
A = f(E_s, E_p, E_a, C),
\]
where \(f(\cdot)\) denotes an evidence-integration process rather than a specific statistical model.

For an audit case \(i\), we operationalize the evidence profile as
\[
E_i=(g_i,p_i,w_i,n_i),
\]
where \(g_i\) is the observed fairness gap, \(p_i\) the estimated probability that the absolute gap exceeds an actionable threshold, \(w_i\) an uncertainty measure such as confidence-interval width, and \(n_i\) a subgroup adequacy measure such as the number of positive-label cases in the relevant subgroup---so that \(g_i,p_i\) capture Signal Evidence, \(w_i\) captures Precision Evidence, and \(n_i\) captures Adequacy Evidence.

We estimate Statistical Actionability as
\[
\hat{A}_i = h(E_i;\theta),
\]
where \(h(\cdot)\) is an evidence-integration function and \(\theta\) denotes context-determined evidence requirements. Different procedures may estimate \(p_i\) or \(w_i\), and different contexts may impose different requirements; the framework fixes only that actionability is inferred from accumulated evidence rather than read directly from the observed gap.

\begin{figure*}[t]
\centering
\includegraphics[width=0.95\textwidth]{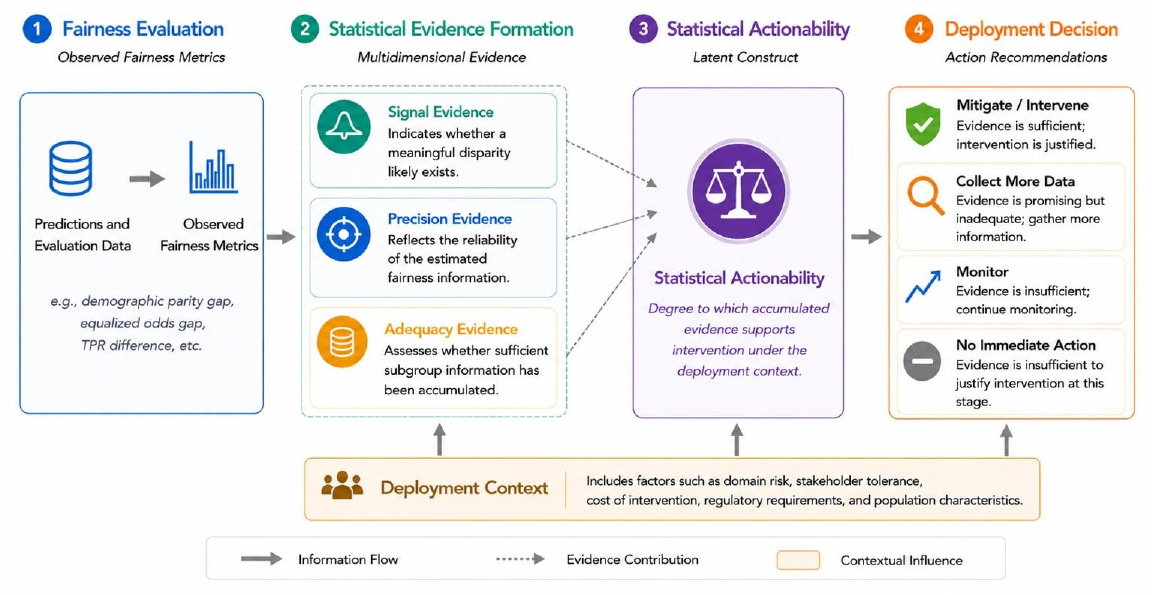}
\caption{Conceptual architecture of Statistical Actionability. Observed fairness metrics serve as inputs to statistical evidence formation rather than direct deployment triggers. Signal, precision, and adequacy evidence are integrated with deployment context to infer Statistical Actionability, which is then translated into a deployment recommendation.}
\label{fig:conceptual_framework}   
\end{figure*}

Figure~\ref{fig:conceptual_framework} summarizes the conceptual architecture.

\subsection{Deployment Decision Rule}
\label{sec:decision-rule}

Statistical Actionability must ultimately be translated into a deployment recommendation \(D_i\), drawn from
\[
\mathcal{D}
=
\{
\mathrm{mitigate},
\mathrm{collect\_data},
\mathrm{monitor},
\mathrm{no\_action}
\},
\]
via a decision rule
\[
D_i = \delta(\hat{A}_i,E_i,C_i;\theta),
\]
where \(C_i\) denotes deployment-context information and \(\theta\) specifies the evidence requirements for action.

This formulation distinguishes estimated actionability from the final recommendation. \(\hat{A}_i\) summarizes overall evidentiary strength, but the recommendation also depends on which evidence dimension is limiting---a potentially meaningful gap with insufficient subgroup support calls for further data collection, while moderate but uncertain evidence calls for monitoring---which is why \(E_i\) enters \(\delta(\cdot)\) directly rather than only through \(\hat{A}_i\). The same observed gap can therefore lead to different recommendations depending on uncertainty, adequacy, and context, and the same estimated actionability may be handled differently when the consequences of acting or waiting differ across settings.

The four recommendations carry distinct statistical meanings. \textbf{Mitigate} indicates evidence sufficiently strong, reliable, and adequate to justify immediate intervention. \textbf{Collect data} indicates a potentially important disparity with currently insufficient evidence. \textbf{Monitor} indicates a potential concern that does not yet justify intervention or targeted data collection. \textbf{No action} indicates that available evidence does not currently support intervention. These categories are used consistently throughout the empirical evaluation; the complete stepwise decision procedure is provided in the supplementary material.

The decision rule does not optimize the observed gap directly; its role is to determine whether the evidence behind that gap clears the bar for intervention. Conventional fairness auditing maps disparities directly to action, whereas the proposed procedure inserts an explicit Statistical Actionability layer between evaluation and action.

\subsection{Stability-based Deployment Reference}
\label{sec:sdr}

Sections~\ref{sec:decision-framing}--\ref{sec:decision-rule} define Statistical Actionability as a construct and translate it into a decision rule. Evaluating that rule requires a different strategy from conventional supervised learning: standard prediction tasks compare outputs against observed labels, but fairness deployment benchmarks provide no label indicating whether a disparity \emph{should} trigger intervention, making classification-style evaluation inapplicable.

We address this with a \textbf{Stability-based Deployment Reference} (SDR), which estimates whether a recommendation remains statistically stable under repeated sampling variation. SDR is an empirical stability reference, not a ground-truth oracle: stability under resampling is not a claim that a recommendation is morally, legally, or normatively correct.

Formally, let \(R_i = r(E_i)\), \(R_i \in \mathcal{D}\), where \(r(\cdot)\) denotes the reference-construction procedure. Operationally, SDR is constructed through repeated evaluation-set resampling: for each audit case, we bootstrap the evaluation data while keeping the trained model fixed, recompute fairness statistics, reconstruct the evidence profile, and apply the reference-construction rule. The resulting empirical distribution over \(B\) bootstrap resamples characterizes how consistently the available evidence supports each action; recommendations stable across resampling are interpreted as having stronger empirical support than those that fluctuate substantially. The full construction procedure and stability criteria are provided in the supplementary material.

\subsection{Representative Statistical Audit Selection}
\label{sec:audit-selection}

Aggregate evaluation metrics summarize overall deployment behavior but do not explain why recommendations change in individual audits. To support transparent interpretation without anecdotal reporting, we select representative audits through a predefined statistical protocol.

Each case produces two recommendations---\(D_i^{\mathrm{gap}}\) from a conventional observed-gap rule and \(D_i^{\mathrm{SAP}}\) from Statistical Actionability---defining a decision transition \(T_i=(D_i^{\mathrm{gap}},D_i^{\mathrm{SAP}})\) such as mitigate-to-mitigate, mitigate-to-collect-data, mitigate-to-monitor, or no-action-to-monitor, representing changes in deployment behavior rather than different fairness outcomes.

For each case, we construct a standardized evidence vector \(x_i=(|g_i|,p_i,w_i,n_i)\)---the absolute gap, actionability probability, uncertainty measure, and subgroup adequacy measure---standardized within each transition category. For category \(T\), with centroid \(\bar{x}_T\) and covariance \(\Sigma_T\) of the standardized vectors, we compute the Mahalanobis distance
\[
d_i=(x_i-\bar{x}_T)^\top \Sigma_T^{-1}(x_i-\bar{x}_T)
\]
and select \(i_T^\ast = \arg\min_{i:T_i=T} d_i\) as the representative audit, using the Moore--Penrose pseudoinverse when \(\Sigma_T\) is singular. Categories with a single case are reported by definition.

This procedure eliminates subjective example selection in favor of a reproducible rule, ensuring reported examples characterize the typical evidence configuration of each transition rather than isolated or extreme observations. The complete selection procedure is provided in the supplementary material.

\section{Results}
\label{sec:results}

\subsection{Overall Effectiveness and Partial-Evidence Baselines}
\label{sec:results-overall}

Table~\ref{tab:overall} compares Statistical Actionability with representative deployment baselines. Two partial-evidence baselines test whether a single evidence dimension is sufficient for deployment decisions. \textbf{CI-only} relies on statistical precision alone, whereas \textbf{Practical-threshold-only} relies only on disparity magnitude.

The gap-only rule frequently overreacted to observed disparities, producing a false-alarm rate of 37.38\% and an average decision cost of 1.195. At the opposite extreme, always mitigating nearly eliminated missed bias, reducing it to 0.46\%, but generated false alarms in 99.54\% of non-actionable cases and incurred a decision cost of 1.505. Statistical Actionability achieved the best overall deployment performance: it reduced false alarms to 3.68\% while maintaining a missed-bias rate of 4.79\%, yielding the lowest decision cost among all evaluated strategies at 0.966.

\begin{table}[t]
\centering
\small
\caption{Performance of deployment decision strategies. FA denotes the false-alarm rate and MB denotes the missed-bias rate. Decision cost denotes the average deployment loss under the cost function defined in Section~\ref{sec:decision-rule}; lower values indicate better deployment performance.}
\label{tab:overall}
\begin{tabular}{lccc}
\hline
Method & FA (\%) & MB (\%) & Cost \\
\hline
Gap-only                 & 37.38 & 16.05 & 1.195 \\
CI-only                  & 22.70 & 44.83 & 1.624 \\
Practical-threshold-only & 29.36 & 30.08 & 1.395 \\
Always mitigate          & 99.54 &  0.46 & 1.505 \\
\textbf{Statistical Actionability}
                         & \textbf{3.68}
                         & \textbf{4.79}
                         & \textbf{0.966} \\
\hline
\end{tabular}
\end{table}

Neither partial-evidence strategy performs well in isolation. CI-only substantially reduced false alarms relative to gap-only, from 37.38\% to 22.70\%, but did so by sharply increasing missed bias to 44.83\%. Precision information without an explicit actionability signal was therefore overly conservative. Practical-threshold-only produced a more balanced error profile, but still incurred a false-alarm rate of 29.36\%, a missed-bias rate of 30.08\%, and a decision cost of 1.395. Notably, CI-only achieved a lower false-alarm rate than Practical-threshold-only but a higher total cost because the cost function assigns greater weight to missed bias. Minimizing either error independently therefore does not minimize overall deployment loss.

Relative to gap-only, Statistical Actionability reduced average decision cost by 19.2\%, from 1.195 to 0.966, and was the only evaluated strategy to substantially reduce both error types simultaneously. The gain therefore arises from jointly integrating disparity magnitude, statistical precision, estimated actionability, and subgroup adequacy rather than from any single criterion. These results validate the central premise of Statistical Actionability: deployment decisions should be driven by accumulated statistical evidence rather than observed fairness gaps alone.

\subsection{Deployment Regions under Statistical Actionability}
\label{sec:results-regions}

Deployment recommendations are not determined by observed fairness gaps alone. Instead, the statistical evidence space is partitioned into decision regions defined jointly by Statistical Actionability and evidence reliability.

\begin{figure*}[t]
\centering
\includegraphics[width=0.95\textwidth]{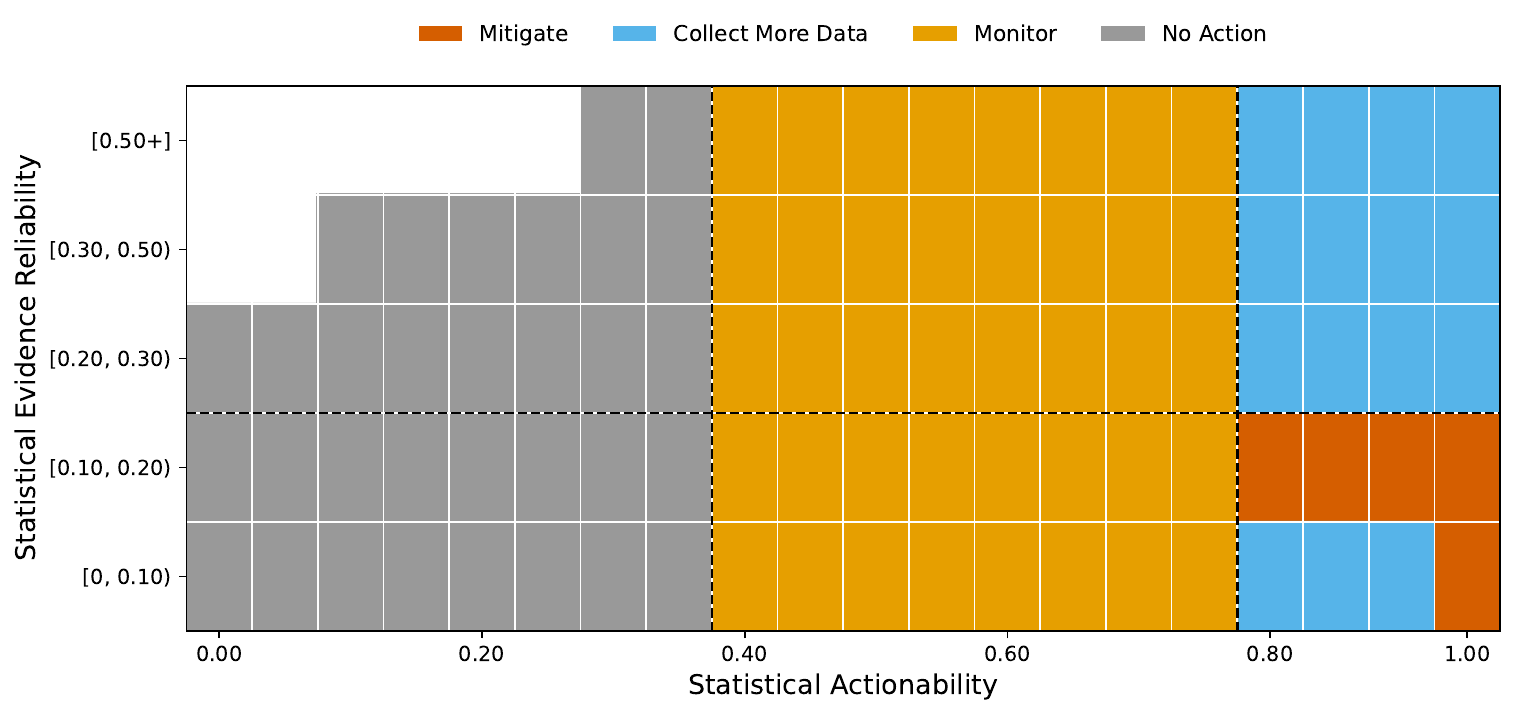}
\caption{Deployment decision map under Statistical Actionability. The evidence space is partitioned into four deployment recommendations according to Statistical Actionability and evidence reliability.}
\label{fig:deployment_map}  
\end{figure*}

Figure~\ref{fig:deployment_map} presents the resulting deployment decision map across the four recommendations defined in Section~\ref{sec:decision-rule}: \textsc{Mitigate}, \textsc{Collect Data}, \textsc{Monitor}, and \textsc{No Action}. These regions represent qualitatively different levels of evidentiary support rather than different ranges of disparity magnitude. Immediate mitigation occupies only a relatively small region of the evidence space and emerges when actionability is high and the supporting evidence is sufficiently reliable. A substantial portion of the space instead corresponds to intermediate responses: potentially actionable but unreliable evidence is assigned to \textsc{Collect Data}, emerging but inconclusive evidence to \textsc{Monitor}, and low-actionability cases to \textsc{No Action}.

Figure~\ref{fig:deployment_map} therefore provides empirical support for the Statistical Actionability Principle by showing that deployment recommendations emerge from the interaction between accumulated actionability and evidentiary reliability rather than from isolated fairness metrics. Fairness evaluation consequently becomes an evidence-based deployment decision problem rather than a threshold-crossing exercise.

\subsection{Transportability across Deployment Regimes}
\label{sec:results-transport}

We next evaluate whether the calibrated deployment rule remains effective when future deployment environments differ from the calibration environment. Five target regimes represent distinct statistical conditions: rare bias (T1), class imbalance (T2), rare outcomes (T3), noisy model behavior (T4), and off-grid fairness gaps (T5).

\begin{figure*}[t]
\centering
\includegraphics[width=0.95\textwidth]{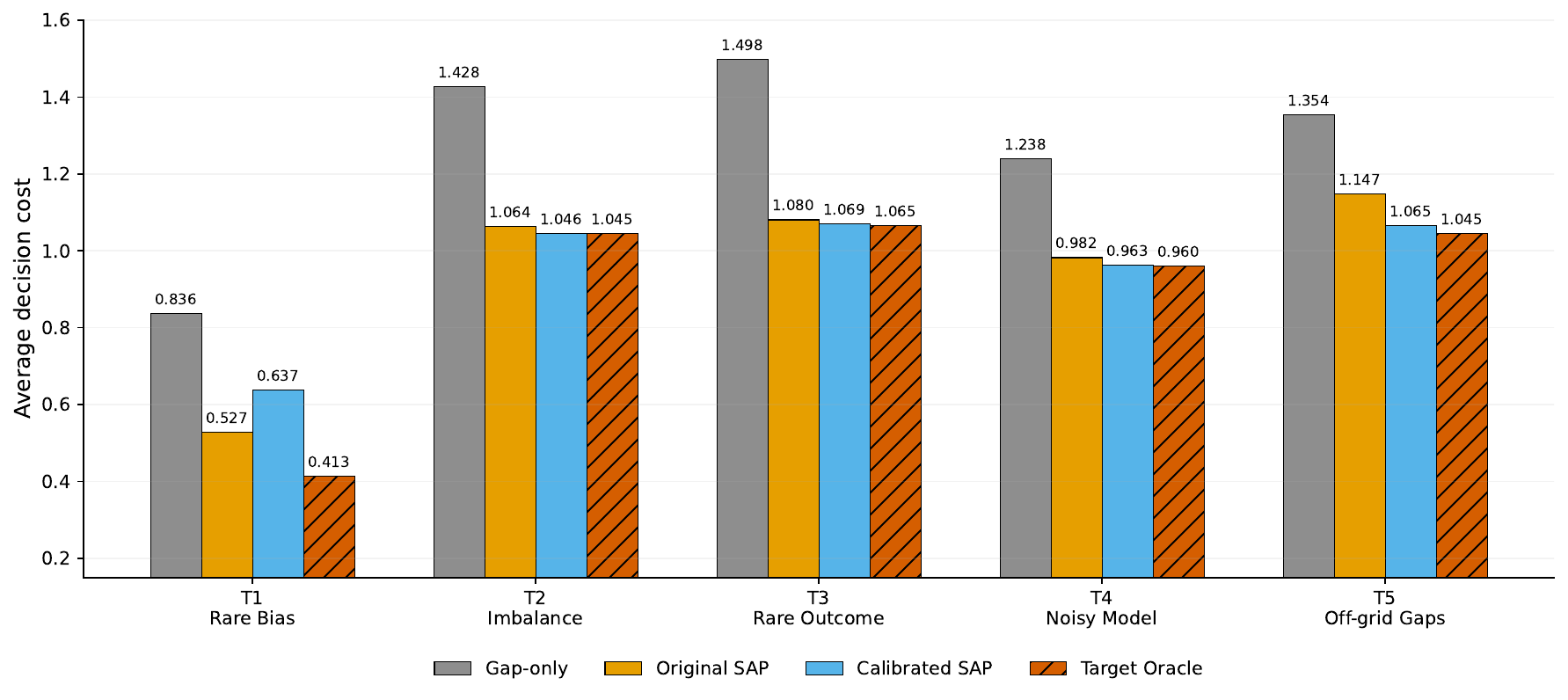}
\caption{Transportability of Statistical Actionability across five target deployment regimes. Bars report average decision cost; lower values indicate better deployment performance. The target oracle provides a regime-specific reference.}
\label{fig:transport}
\end{figure*}

Figure~\ref{fig:transport} shows that both SAP variants achieved lower decision cost than gap-only in all five regimes. The calibrated SAP improved on the original SAP in four of the five regimes and remained close to the target oracle in T2--T5. Table~\ref{tab:transport} quantifies this transportability. The calibrated SAP remained within 2\% of the target oracle in T2--T5, indicating that one calibrated rule closely approximated target-specific oracle references across four substantially different environments without regime-specific recalibration.

\begin{table}[t]
\centering
\small
\caption{Relative decision-cost gap to the target oracle. Smaller values indicate better transportability.}
\label{tab:transport}
\begin{tabular}{lcc}
\hline
Target regime & Original SAP (\%) & Calibrated SAP (\%) \\
\hline
T1 Rare Bias     & 27.80 & 54.50 \\
T2 Imbalance     &  1.82 & \textbf{0.07} \\
T3 Rare Outcome  &  1.45 & \textbf{0.40} \\
T4 Noisy Model   &  2.32 & \textbf{0.34} \\
T5 Off-grid Gaps &  9.82 & \textbf{1.98} \\
\hline
\end{tabular}
\end{table}

T1 was the only clear exception. The calibrated SAP incurred a decision cost of 0.637, compared with 0.527 for the original SAP and 0.413 for the target oracle, corresponding to a 54.5\% relative oracle gap. Under extremely rare bias, very limited statistical evidence accumulates for the underlying disparity, causing the calibrated rule to become overly conservative and delay intervention. We treat this behavior as a boundary condition of the current framework and return to it in Section~\ref{sec:discussion}.

Overall, the transportability results suggest that Statistical Actionability captures evidence-accumulation patterns that generalize across heterogeneous deployment environments rather than merely fitting a particular calibration regime. The calibrated rule remained close to the target oracle in four of five regimes while retaining lower decision cost than gap-only throughout.

\subsection{Real-World Deployment Characterization}
\label{sec:results-realworld}

\begin{figure*}[t]
\centering
\includegraphics[width=0.95\textwidth]{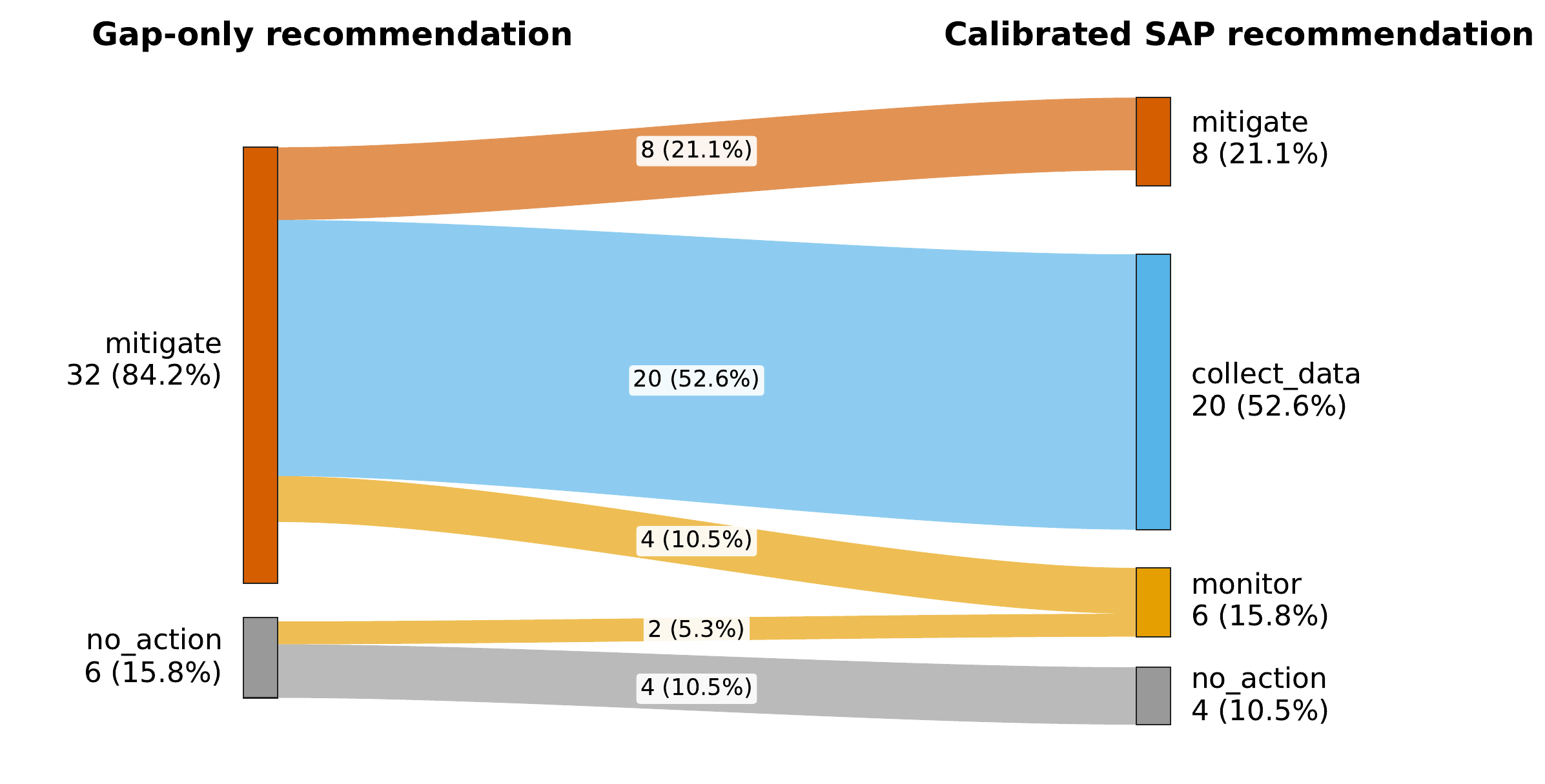}
\caption{Decision transitions from gap-only recommendations to calibrated Statistical Actionability recommendations. Flow width represents the number of statistical audits assigned to each transition.}
\label{fig:transitions}
\end{figure*}

\begin{figure*}[t]
\centering
\includegraphics[width=0.95\textwidth]{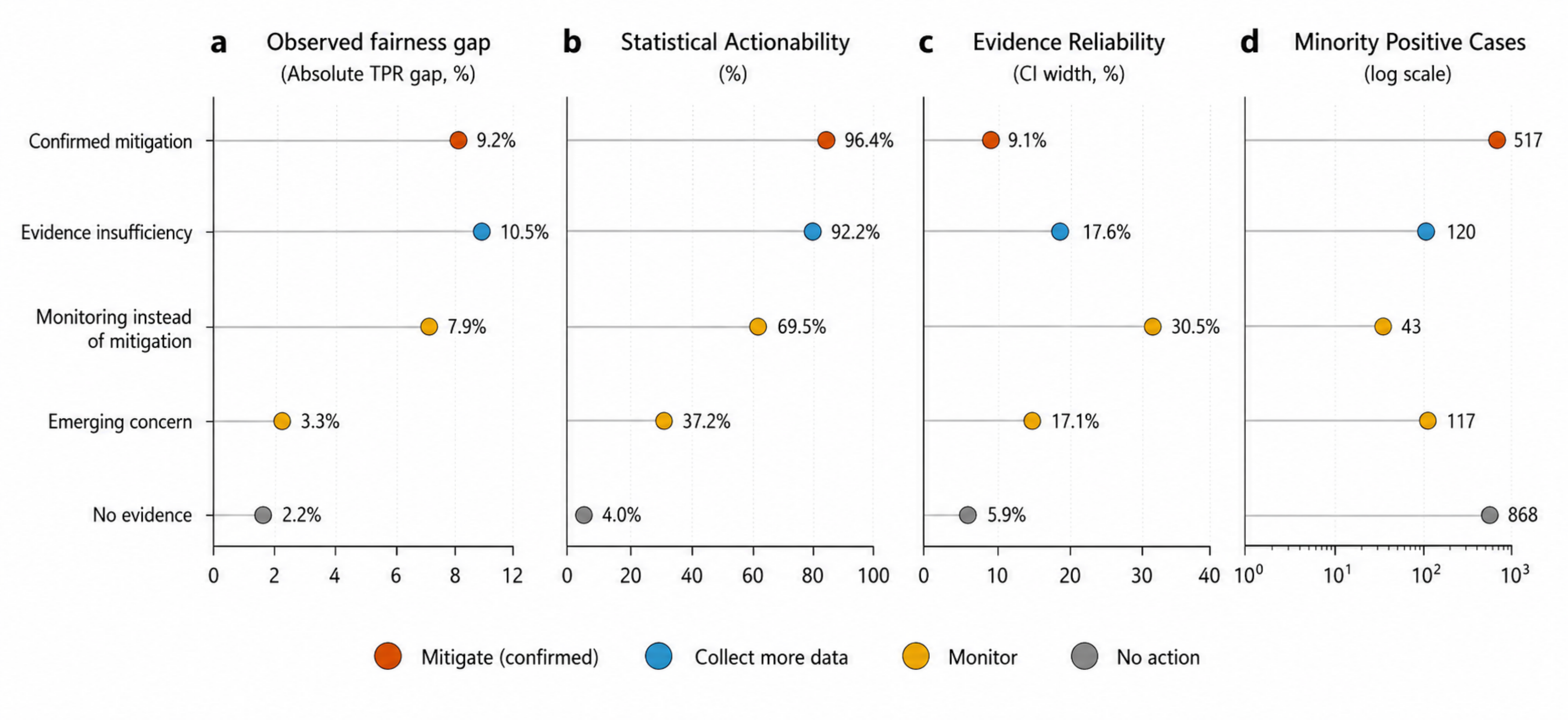}
\caption{Representative statistical audit profiles across major decision-transition categories. Profiles are characterized by observed fairness gap, Statistical Actionability, confidence-interval width, and minority positive-case count.}
\label{fig:profiles}
\end{figure*}

We finally examine how Statistical Actionability changes recommendations in real-world fairness audits and how the resulting actions can be interpreted through representative statistical profiles. Figure~\ref{fig:transitions} summarizes transitions from gap-only to calibrated-SAP recommendations, while Figure~\ref{fig:profiles} characterizes the evidence patterns underlying the major transition categories.


\textbf{Decision transitions.}

The gap-only rule recommended mitigation in 32 of 38 audits, corresponding to 84.2\% of cases. After incorporating accumulated statistical evidence, only 8 audits, or 21.1\%, remained immediate-mitigation recommendations. The largest transition was from \textsc{Mitigate} to \textsc{Collect Data}: 20 audits, representing 52.6\% of all cases, were redirected to evidence acquisition because the available evidence was insufficiently reliable for immediate intervention. Four audits, or 10.5\%, moved from \textsc{Mitigate} to \textsc{Monitor}, reflecting concerns that did not meet the evidentiary requirement for immediate action. Among the six audits initially assigned \textsc{No Action}, four remained unchanged and two transitioned to \textsc{Monitor}.

The dominant transition was therefore from mitigation to evidence acquisition rather than from mitigation to no action. Statistical Actionability primarily postpones intervention until sufficient evidence accumulates, reallocating deployment actions according to evidence strength and reliability rather than simply reducing mitigation frequency.

\textbf{Representative statistical audit profiles.}
Figure~\ref{fig:profiles} summarizes one representative audit for each major transition category, selected by proximity to the center of its standardized evidence distribution as described in Section~\ref{sec:audit-selection}. Each profile is characterized along four dimensions: observed fairness gap, Statistical Actionability, confidence-interval width, and minority positive-case count.

The confirmed-mitigation and evidence-insufficiency profiles have comparable observed gaps but receive different recommendations. Confirmed mitigation combines high Statistical Actionability with a narrow confidence interval and a comparatively large minority-positive sample. Evidence insufficiency also exhibits high estimated actionability, but its wider confidence interval and substantially smaller subgroup sample lead the framework to recommend additional data collection rather than immediate mitigation.

The monitoring-instead-of-mitigation profile exhibits moderate Statistical Actionability, the widest confidence interval among the selected audits, and a small minority-positive sample, indicating a potentially consequential but highly uncertain disparity. The emerging-concern profile has lower actionability but non-negligible evidence, supporting continued monitoring. The no-evidence profile combines a small observed gap and low actionability with a large subgroup sample, warranting no immediate action.

Together, these profiles show that comparable observed disparities can lead to different deployment recommendations depending on evidence reliability, subgroup adequacy, and accumulated Statistical Actionability. Figure~\ref{fig:profiles} therefore provides an interpretable statistical account of why the calibrated SAP treats apparently similar audits differently.

Collectively, these experiments demonstrate that Statistical Actionability improves deployment quality, distinguishes substantively different evidence states, generalizes across heterogeneous deployment environments, and produces recommendations that remain traceable to their underlying statistical evidence.

\section{Discussion and Conclusion}
\label{sec:discussion}

\subsection{Statistical Actionability as a Statistical Construct}

The principal contribution of this work is not a new fairness metric or deployment algorithm, but the formalization of \emph{Statistical Actionability} as a statistical construct for responsible AI deployment. Existing fairness research has substantially advanced the measurement of algorithmic disparities, uncertainty estimation, and fairness-aware optimization. These approaches primarily address whether disparities exist, how reliably they can be estimated, or how model behavior can be modified. They do not directly answer the deployment question practitioners face: \emph{When is the available statistical evidence sufficient to justify intervention?}

We argue that Statistical Actionability should be regarded as a statistical construct in the same broad sense as calibration and uncertainty quantification. None prescribes a single estimation algorithm; instead, each characterizes a distinct statistical property---predictive reliability, predictive uncertainty, or evidentiary sufficiency for action---that may be estimated through multiple methodologies. Platt scaling, isotonic regression, and temperature scaling are procedures for estimating or improving calibration, but they are not themselves synonymous with calibration. Similarly, different inferential procedures, Bayesian formulations, or evidence-accumulation strategies may operationalize Statistical Actionability without defining the construct itself. The underlying object of interest remains whether the available evidence is sufficient to support intervention.

This distinction positions Statistical Actionability as a property of the available evidence rather than of any particular classifier, fairness metric, or deployment algorithm. It also separates the construct from the specific operationalization examined in this paper. The posterior evidence measure, confidence-interval requirement, and subgroup-adequacy gate provide one implementation of Statistical Actionability, but alternative estimators may instantiate the same construct under different modeling assumptions or deployment contexts.

More broadly, the framework changes the role of fairness evaluation. Observed disparities are no longer treated as direct deployment triggers; instead, they become one component of an evidence-accumulation process that integrates disparity magnitude, statistical precision, subgroup adequacy, and deployment context. Revisiting Table~\ref{tab:positioning}, Statistical Actionability addresses the decision layer left insufficiently formalized across fairness measurement, uncertainty-aware evaluation, and operational fairness research. It is through this layer that accumulated evidence, rather than an isolated fairness measurement, comes to justify deployment action.

\subsection{Implications for Responsible AI Deployment}

The proposed framework has several implications for responsible AI deployment. Most fundamentally, fairness auditing should be understood as an evidence-based decision problem rather than a threshold-detection exercise. As the empirical results demonstrate, fairness gaps of similar magnitude can correspond to substantially different levels of statistical support. Deployment action should therefore depend on accumulated evidence rather than on observed disparity alone.

This reframing also broadens the set of legitimate deployment responses. Conventional workflows often reduce fairness decisions to a binary choice between mitigating bias and taking no action. Statistical Actionability instead distinguishes among multiple evidence states that warrant different operational responses. Collecting additional data is appropriate when a potentially important disparity is accompanied by inadequate precision or subgroup support. Monitoring is appropriate when evidence is emerging but remains insufficient for intervention. These responses are not failures to act; they are actions directed toward resolving evidentiary uncertainty. Responsible deployment is therefore better understood as an iterative process of evidence accumulation, reassessment, and proportionate intervention.

The framework further clarifies the relationship between statistical evidence and organizational judgment. Statistical Actionability does not eliminate normative or contextual decision making. The practical importance of a disparity, the relative costs of missed bias and unnecessary intervention, and the acceptable evidentiary standard remain deployment-specific. The framework instead makes those requirements explicit and separates them from the observed fairness metric. This separation can improve audit transparency by allowing stakeholders to identify whether disagreement arises from the statistical evidence, the operational cost structure, or the normative standard applied to intervention.

Finally, Statistical Actionability offers a statistical mechanism compatible with broader principles of responsible AI governance. Risk-based oversight increasingly emphasizes proportional responses, documented evidence, and continuous monitoring over the deployment lifecycle. The proposed framework operationalizes these principles by translating heterogeneous statistical evidence into interpretable recommendations. It does not replace existing fairness metrics; it provides the statistical bridge connecting fairness measurement to deployment action.

\subsection{Boundary Conditions of Statistical Actionability}

Every statistical construct has a domain of validity, and Statistical Actionability is no exception. The transportability analysis identifies a clear boundary condition under extremely rare actionable bias, although other extreme regimes may produce related effects.

When actionable fairness violations become exceedingly rare, informative observations accumulate slowly. Consequently, a larger amount of data is required to attain the same level of evidentiary reliability. Fixed intervention requirements therefore become more difficult to satisfy, and deployment recommendations become increasingly conservative. This mechanism explains why the calibrated rule transported less effectively in T1 than in the remaining regimes.

This result does not contradict the Statistical Actionability Principle. On the contrary, it follows directly from its evidence-based logic. A framework designed to recommend intervention only after sufficient evidence has accumulated should delay action when the available data cannot reliably distinguish a persistent disparity from sampling variation. The resulting conservatism reflects an information constraint, not a unique defect of the proposed method.

Comparable boundary conditions arise throughout statistical practice. Calibration assessment becomes more difficult in severely imbalanced settings, and uncertainty estimates become less informative under sparse observation. Statistical Actionability faces an analogous limitation: when decision-relevant evidence is intrinsically scarce, reliable actionability cannot be produced without additional assumptions, information, or data. The decline in transportability therefore identifies the applicability boundary of the current operationalization rather than a failure of the underlying construct.

This boundary has an important practical implication. Rare but high-consequence fairness violations may require evidence strategies different from those used in routine auditing. External data pooling, hierarchical modeling, informative priors, targeted subgroup sampling, or precautionary policies may be needed when waiting for conventional evidence accumulation would impose unacceptable social costs. Statistical Actionability clarifies why such departures are necessary: the deployment context changes the evidence that can reasonably be required before action.

\subsection{Limitations and Future Research}

Several limitations define priorities for future research. Although the evaluation combines controlled simulations with real-world benchmark audits, it does not observe longitudinal decisions from operational AI deployments. The benchmark analysis demonstrates how Statistical Actionability reshapes recommendations under realistic data structures, but it cannot capture organizational responses, changing stakeholder preferences, feedback effects, or the consequences of interventions over time. Future work should evaluate the construct using repeated fairness audits from deployed systems and examine whether evidence-based recommendations improve actual governance outcomes.

The current operationalization also uses a limited set of evidence sources: observed disparities, posterior evidence, confidence-interval width, subgroup adequacy, and deployment costs. These dimensions provide a statistically coherent starting point, but they do not exhaust the evidence relevant to responsible deployment. Causal evidence, distribution shift, temporal fairness drift, human-auditor assessments, measurement validity, and domain-specific harms may alter whether intervention is justified. Future research should investigate how heterogeneous and potentially conflicting evidence can be integrated without obscuring the interpretation of Statistical Actionability.

A further limitation concerns the mapping from evidence to action. The present implementation uses calibrated thresholds to produce four discrete recommendations. This structure improves interpretability but treats the deployment policy as static. In practice, evidence evolves, intervention costs change, and previous actions affect subsequent data. Bayesian updating, sequential decision theory, online calibration, and adaptive policy learning could support actionability estimates that change as new evidence arrives. Such methods should preserve the separation between the construct and its estimator: an adaptive algorithm would estimate Statistical Actionability, not redefine it.

Future work should also develop the formal statistical theory of the construct. Important questions include identifiability, consistency, calibration of actionability estimates, finite-sample guarantees, sensitivity to cost misspecification, and transportability under distribution shift. Domain-specific versions may require different evidence requirements while retaining a shared theoretical core. Just as calibration and uncertainty quantification have developed into broad methodological research programs, Statistical Actionability may support multiple estimators, diagnostics, guarantees, and application-specific formulations.

\bigskip

\noindent\textbf{Conclusion.}
This paper formalizes Statistical Actionability as a statistical construct for evidence-based fairness deployment. Rather than treating observed disparities as direct intervention triggers, the proposed framework integrates disparity magnitude, statistical uncertainty, subgroup adequacy, and deployment context to determine whether intervention is statistically justified. Across controlled simulations, Statistical Actionability reduced deployment decision cost and transported across most evaluated regimes. Across real-world benchmark audits, it produced interpretable decision transitions and evidence profiles that clarified why superficially similar fairness gaps warranted different deployment responses. These findings establish Statistical Actionability as a promising statistical object for connecting fairness measurement to responsible deployment action, analogous to the roles that calibration and uncertainty quantification play in predictive modeling.

\begingroup
\small
\setlength{\bibsep}{3pt plus 1pt minus 1pt}
\bibliographystyle{abbrvnat}
\bibliography{bib}
\endgroup
\end{document}